# Field Evaluation of Column CO$_2$ Retrievals from Intensity-Modulated Continuous-Wave Differential Absorption Lidar Measurements during ACT-America

Joel F. Campbell[1], Bing Lin[1], Michael D. Obland[1], Jeremy Dobler[3], Wayne Erxleben[4], Doug McGregor[3], Chris O'Dell[4], Emily Bell[4], Sandip Pal[5], Brad Weir[6,10], Tai-Fang Fan[7], Susan Kooi[7], Abigail Corbett[7], Kenneth Davis[8], Iouli Gordon[9], Roman Kochanov[11]

[1]NASA Langley Research Center (LaRC), [2]Spectral Sensor Solutions LLC, [3]L3Harris Technologies, [4]Colorado State University, [5]Department of Geosciences, Atmospheric Science Division, Texas Tech University, [6]Universities Space Research Association, [7]Science System and Application, Inc, [8]The Pennsylvania State University, [9]Harvard-Smithsonian Center for Astrophysics, [10]NASA Goddard Space Flight Center, [11]Laboratory of Theoretical Spectroscopy, V.E. Zuev Institute of Atmospheric Optics, Tomsk, Russia

Abstract

We present an evaluation of airborne Intensity-Modulated Continuous-Wave (IM-CW) lidar measurements of atmospheric column CO$_2$ mole fractions during the ACT-America project. This lidar system transmits online and offline wavelengths simultaneously on the 1.57111-µm CO$_2$ absorption line, with each modulated wavelength using orthogonal swept frequency waveforms. After the spectral characteristics of this system were calibrated through short-path measurements, we used the HITRAN spectroscopic database to derive the average-column CO$_2$ mixing ratio (XCO$_2$) from the lidar measured optical depths. Based on in situ measurements of meteorological parameters and CO$_2$ concentrations for calibration data, we demonstrate that our lidar CO$_2$ measurements were consistent from season to season and had an absolute calibration error (standard deviation) of 0.80 ppm when compared to XCO$_2$ values derived from in situ measurements. By using a 10-second or longer moving average, a long-term stability of 1 ppm or better was obtained. The estimated CO$_2$ measurement precision for 0.1-s, 1-s, 10-s, and 60-s averages were determined to be **3.4 ppm (0.84%), 1.2 ppm (0.30%), 0.43 ppm (0.10%), and 0.26 ppm (0.063%), respectively.** These correspond to measurement signal-to-noise ratios of **120, 330, 950, and 1600**, respectively. The drift in XCO$_2$ over one-hour of flight time was found to be below our detection limit of about 0.1 ppm. These analyses demonstrate that the measurement stability, precision and accuracy are all well below the thresholds needed to study synoptic-scale variations in atmospheric XCO$_2$.

## 1. Introduction

Atmospheric carbon dioxide (CO$_2$) is a crucial part of the Earth's carbon cycle and one of the major greenhouse gases (GHGs) in the Earth's climate system. The CO$_2$ concentration within the atmosphere has significantly changed over the last 150 years, due mainly to anthropogenic activities. Understanding the carbon cycle is essential for diagnosing current, and predicting future, climate change (Marquis and Tans, 2008; Gregory et al., 2009; Michalak et al., 2011). The Earth's terrestrial biosphere has been a strong net sink of atmospheric CO$_2$ for decades (e.g., LeQuere et al., 2009), substantially slowing the rate of accumulation of CO$_2$ in the atmosphere from combustion of fossil fuels. The causes of the net biogenic $CO_2$ sink, its location and magnitude (Peylin et al., 2013), and its likely evolution in the future (e.g., Friedlingstein et al., 2006) all remain highly uncertain, contributing substantial uncertainty to the projections of future climate (Stocker et al., 2013). North American biogenic $CO_2$ fluxes, for example, are estimated on a 5-year, continentally-aggregated basis to an accuracy no better than 50% (SOCCR, 2007; King et al., 2012). Individual annual estimates from biosphere models, biomass inventories, and atmospheric inversions (Hayes et al., 2012; Peylin et al., 2013) also often diverge by a factor of 2. Uncertainties in $CO_2$ transport estimates may contribute a significant part of the flux uncertainties. With constraints from satellite column $CO_2$ observations, the flux uncertainties can be reduced (Lauvaux et al., 2012; Liu et al., 2017; Crowell et al., 2019).

Comprehensive measurements of regional atmospheric $CO_2$ distributions are urgently needed to develop a more complete understanding of $CO_2$ transport and sources and sinks. This is one of the key reasons NASA has been carrying out the Atmospheric Carbon and Transport–America (ACT–America) suborbital mission. Another ACT-America objective is to compare the full-column $CO_2$ spatial variability observed by the NASA Orbiting Carbon Observatory – 2 (OCO–2) satellite mission with the partial-column $CO_2$ lidar measurements obtained below the ACT-America C-130 aircraft across the lower troposphere.

The target areas of the ACT–America mission are the mid-Atlantic, mid-west, and Gulf-coast regions of the United States. The mission has completed four deployments with $CO_2$ lidar: one for each season (summer 2016; winter 2017; fall 2017; spring 2018) in each of these areas, as well as a fifth deployment without $CO_2$ lidar measuring summer a second time in 2019 due to the strong atmosphere-biosphere interaction during this season. Two NASA aircraft (C-130 and B-200) were deployed during each field campaign. Both aircraft were equipped with GPS for geographic location, altitude, and time information and in situ sensors for the measurement of atmospheric $CO_2$, other greenhouse gases, and meteorological variables such as temperature, pressure, humidity, and wind. Besides these in situ sensors, the C-130 also carried remote sensors including the Multi-Functional Fiber Laser Lidar (MFLL; Dobler et al., 2013), which is the core instrument for atmospheric column $CO_2$ measurements beneath the aircraft, and the Cloud Physics Lidar (CPL), an airborne lidar system designed specifically for studying clouds, aerosols and atmospheric boundary layer heights (McGill et al., 2002).

MFLL is an intensity-modulated continuous-wave (IM-CW) lidar developed by the Harris, Corp. and the NASA Langley Research Center for column $CO_2$ measurements. The Integrated Path Differential Absorption (IPDA) technique is used to measure column $CO_2$ differential absorption optical depth (DAOD) based on the combined lidar measurements at online and offline wavelengths using the $CO_2$ absorption line at 1.57111 μm (Browell et al., 2008; Dobler et al., 2013) and this corresponds to the peak center at 15,000 ft altitude. The offlines were +/- 50 pm from the centerline. NASA Langley Research Center and the Harris, Corp. have jointly demonstrated its capability for a future space $CO_2$ lidar mission (Browell et al., 2009, 2010a,b; Dobler et al., 2013; Lin et al., 2013). Previous test results from airborne technology demonstration flights were very encouraging (Browell et al., 2009, 2010b; Dobler et al. 2013, Lin et al., 2015). The signal-to-noise ratio (SNR) for clear sky IPDA measurements of $CO_2$ DAOD for a 10-s average over vegetated areas from a 7-km altitude was found to be as high as 1300, resulting in an estimated precision uncertainty of 0.077% or an equivalent $XCO_2$ precision of ~0.3 ppm. A critical part of high precision $XCO_2$ measurements is the ranging capability, which is realized by the use of advanced waveform modulation and super-resolution techniques (Campbell et al., 2014a, b). The range uncertainty has been demonstrated to be below the sub-meter level (Campbell et al., 2014a, b). With this ranging capability, $XCO_2$ values can also be retrieved for the column between the aircraft and optically thick cloud decks and for weather conditions with intervening thin cirrus clouds or other aerosol layers (Lin et al., 2015).

The extensive ACT-America data set provides further opportunities for MFLL instrument evaluation and applications to carbon cycle science. For atmospheric $CO_2$ transport and flux estimations, MFLL measurements require both high precision and low systematic error. To reach this goal, dedicated MFLL calibration flights under clear to scattered cloud conditions were conducted and post-flight data analyses were performed. This paper discusses the specific MFLL data reduction algorithms used during ACT-America and the resulting DAOD and $XCO_2$ retrievals. Generally, the data reduction covers various aspects of the airborne lidar data processing including consideration of lidar pointing geometry, data quality evaluation, cloudy/clear sky screening, demodulation with a matched filter, and uses of ancillary data such as aircraft time, location and attitude records. The instrument and measurement approach are discussed in the next section. Experiments and calculation procedures associated with these data and data processing are also discussed. Section 3 contains calibration and MFLL measurements under different atmospheric conditions, and an evaluation of MFLL measurement performance, including precision and accuracy during ACT-America. We conclude our main findings in section 4.

## 2. IM-CW Measurement Concept and MFLL Characteristics
### 2.1 Column $CO_2$ measurement concept and instrumentation

The $CO_2$ lidar system, MFLL, deployed during four ACT-America field campaigns has been extensively discussed previously (Browell et al., 2008, 2009, 2010a, b; Dobler et al. 2013; Lin et al., 2013, 2015; Campbell et al. 2014a, b, c, d). Only a brief review is given here for the reader's convenience. The MFLL system uses the IM-CW approach and is built with an all-fiber transmitter. It has relatively low bandwidth, and low peak power requirements, and common-mode noise rejection for all channels from many noise sources, including amplification, atmosphere, ground reflectivity, and detector chain due to simultaneous transmission and reception of all the wavelengths.

The IM-CW method encodes multiple wavelengths with unique amplitude modulation waveforms. The current instrument actually transmits three wavelengths (one online and two offlines on both sides of online) but only two are actually used in the processing. The signals are then combined before amplification and transmission to the atmosphere. The return signals are collected simultaneously on a single detector and digitized. The individual wavelength amplitudes are separated in the digital domain through correlation with a matched filter of the unique encoding waveforms.

The concept of our IM-CW lidar system designed for $CO_2$ IPDA measurements is illustrated in Fig. 1. For the system described in this figure, two intensity modulated seed lasers with their distinct online and offline spectral properties, operating in conjunction with a near infrared $CO_2$ absorption line with one at the online absorption wavelength and the other at an offline absorption wavelength are combined using fiber optics and simultaneously amplified by a single Erbium Doped Fiber Amplifier (EDFA) to increase the transmitted power. A small fraction of the transmitted beam is picked off via an optical tap inside of the EDFA and sent to a reference detector for power normalization.

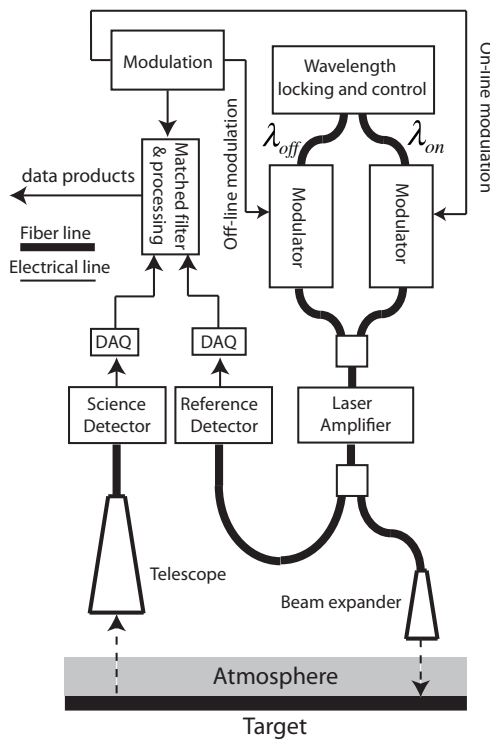

Fig. 1. Baseline instrument block diagram.

This is necessary for accurate IPDA measurements because power fluctuations of the laser system would cause significant errors in $XCO_2$ retrievals. The backscattered science signals of the online and offline wavelengths from the surface as well as aerosols and clouds are simultaneously collected with a telescope, optically filtered with a narrow band optical filter, and detected by the science detector. Both the science and reference signals are amplified, electronically filtered and then digitized for retrievals of column $CO_2$ using IPDA approach. Post processing of the digitized science and reference data allows for discrimination between ground and intermediate scatterers using the matched filter technique. One also obtains differential absorption power ratios for inference of column $CO_2$ amounts as well as range estimates to the scattering targets.

In the type of CW lidar discussed here, a wide band (the bandwidth of the modulation is 500 kHz) modulation signal is used to modulate the intensity of a seed laser, which is then amplified and transmitted through the atmosphere (Fig. 1). The modulation depth is approximately 90%. The digitized received signal is converted into a pulse through a mathematical transformation in the form of a correlation with a complex quadrature reference template stored in memory. The narrowness of the resulting pulse after the correlation is inversely proportional to the bandwidth of the modulation. Different types of modulation produce different pulse shapes or autocorrelation functions. One of the simplest modulations is linear swept frequency (chirp) and that is what is used here. For effective IM-CW performance, the modulation waveforms must exhibit mutual orthogonality (to avoid crosstalk in the demodulation process) and a long enough period to ensure that cloud returns and ground returns can be distinguished unambiguously. However, the type of optical amplifier used in MFLL's transmitter only functions optimally when modulated above about 50 kHz. MFLL currently uses chirps whose instantaneous frequencies begin around 100 kHz and sweep up to about 600 kHz in 200 $\mu$sec, a period that provides an unambiguous range of about 30 km. The number of sweeps per frame is 500, which provides a 10-Hz reporting rate of DAOD measurements. With the 4-MHz signal sampling rate and long-swept-frequency waveforms, significant computational power is needed for both high-precision DAOD and high-resolution range retrievals. A super-resolution FFT interpolation technique is used in data processing. Details of creating this orthogonal modulation and others along with the super-resolution and interpolation techniques can be found in previous studies (Campbell, 2013, Campbell et al., 2014a, b, c, d). A convenient method for determining this correlation is to use

$$R(ref, data) = \frac{1}{N}\sum_{m=0}^{N-1} ref^*(m)\, data(m+n),$$
$$= DFT^{-1}\left(DFT^*(ref^*)DFT(data)\right)$$
(1)

where ref is the reference waveform, data is the data collected either from the reference or science detector, and DFT is the digital Fourier transform. Once this is done on the transmit and receive channels, one may find the one-way DAOD by,

$$\tau_{meas} = \frac{1}{2} \ln\left(\frac{P_{off}^R P_{on}^T}{P_{on}^R P_{off}^T}\right), \tag{2}$$

where $P_{off}^R$ is the received offline power, $P_{on}^R$ is the received online power, $P_{off}^T$ is the transmitted offline power, and $P_{on}^T$ is the transmitted online power.

Although the general architecture remains the same, significant changes to the MFLL system have been implemented since the publication by Dobler et al. (2013). Changes include the consolidation and synchronization of all computer functions, the separation of the single temperature-controlled chamber into three separately controlled chambers, and a microelectromechanical (MEMS) multiplexer to allow a Bristol 621A wavemeter to monitor all produced wavelengths. This has an accuracy of 0.2 pm and these measurements are used to correct for wavelength drift in post processing. Additionally we used a locked laser as a reference to track any potential changes in the wavemeter. For this mission, the MFLL with 5 W transmission power was installed in the NASA C-130 aircraft from the NASA Wallops Flight Facility. Its transmitter and telescope were pointed downward through a custom fused-silica window in the bottom of the fuselage. Note that significant degradation in the window coating was found during the winter 2017 field campaign, resulting in a replacement of the window, which didn't appear to have an impact on data calibration discussed in later sections.

The MFLL uses a simple custom telescope with a parabolic reflector coupled directly into a multi-mode fiber, and housed in a carbon-fiber tube to limit thermal variability, and it is mounted with an approximate pitch offset of 3 degrees relative to the aircraft to compensate for the average aircraft pitch during flight.

## 2.2 Instrument zero path calibration and optical depth offset subtraction

The MFLL calibration approach for compensation of the wavelength-dependent throughput of the internal optics utilizes a very near-field target with the assumption that the absorption over this very short path length is negligible and the power ratio of all channels should be unity after taking this zero-path calibration into consideration. In addition, this calibration accounts for differences in lidar signal path lengths within and outside the instrument such as differences in optical fiber length and the optical path from the telescope to the ground, which is important for correct range measurements, as well as correcting for optical or electrical frequency dependencies of the instrument that could impact the differential absorption measurement. We currently find the distance offset using the offline science channel distance minus the offline reference distance. That distance on the ground is the distance offset for zero path. We also measure the optical depth using Eq. 2 on the ground that becomes our optical depth offset subtraction value. During the ACT-America campaigns, these short path tests were conducted during ground testing before each flight.

## 2.3 Ancillary data for XCO$_2$ Retrieval from DAOD

Atmospheric $CO_2$ absorption is affected by the local meteorological state, including pressure, temperature, and humidity. Thus, meteorological conditions are needed in retrieving XCO$_2$ from the MFLL DAOD measurements. During ACT-America flights, atmospheric $CO_2$ and meteorological profiles were measured by in situ instruments on the C-130 and B-200 (Baier et al, 2020; Pal et al., 2020). In situ profiles measured during spiral segments of ACT-America flights were extremely useful for the calibration and validation of MFLL DAOD measurements; however, there were only a limited number of these profiles available. To derive XCO$_2$ values for all MFLL measured column $CO_2$ DAODs throughout the campaign flights, meteorological profiles along the lidar measurement tracks were needed, as they are for the OCO-2 satellite measurements (O'Dell et al., 2012) and, of course, in situ data is not available at all locations and altitudes. For this study Modern-Era Retrospective analysis for Research and Applications, Version 2 (MERRA-2) reanalysis data (Gelaro et al., 2017; Bosilivich et al., 2017) were therefore used to derive $XCO_2$.

For the purpose of retrieving $XCO_2$ values from the MFLL DAOD measurements at a specific location and time, atmospheric $CO_2$ and water vapor absorption cross-section profiles were calculated at the MERRA-2 grid points and times based on their meteorological profiles including height, temperature, pressure, and humidity and with an assumed atmospheric constant $CO_2$ mixing ratio (e.g., 400 ppm) and a spectroscopic model. The model estimated absorption cross-section profiles were then spatiotemporally interpolated using cubic spline to MFLL measurement track locations and times. These interpolated absorption profiles were then integrated over pressure from the surface to the aircraft altitude to obtain the model-equivalent optical depth for the 400-ppm $XCO_2$ assumption. By scaling the MFLL measured $CO_2$ DAOD with the modeled DAOD for $XCO_2$ of 400 ppm, the MFLL-derived $XCO_2$ value was retrieved The model estimated water vapor DAOD, which is normally very small, was subtracted from the MFLL-measured DAOD to derive the measured $CO_2$ DAOD before the scaling was done.

## 2.4 Range Measurements

The importance of ranging measurements in model DAOD and $XCO_2$ retrievals was mentioned previously. The MFLL native vertical sampling resolution is 37.5 meters but through special techniques that take advantage of the high number of sweeps per frame of data (500 in this case), this resolution is improved by a factor of 500. This resolution improvement is accomplished using a special fine interpolation and super resolution technique that makes a transformation in the frequency domain of the matched filter correlation in real time (Campbell et al, 2014a). Once this is done and transformed back in to the spatial domain, the peak location of the science and reference channels are recorded. To get the final range the reference range is subtracted from the science range and calibration offsets are applied.

## 2.5 Attitude correction

In retrieving the airborne measurements, certain data reduction procedures are applied. The first data screen is aircraft attitude. We only use data with pitch/roll angles less than 5°. This limits the off-nadir pointing to a maximum of ~ 750 m at the surface for a high-altitude flight at ~8.5 km. The retrieved nadir DAOD is estimated from the measured DAOD times a correction. This correction is estimated by finding the vertical component of the unit direction vector of the laser in the earth coordinate system. This is accomplished using an Euler angle transformation that takes into account the orientation of the MFLL with respect to the aircraft inertial navigation unit (INU), in addition to the rotation order of the INU.

## 2.6 Cloud Screening

Besides aircraft attitude correction and range measurements, cloud detection and screening are also performed. Because we need to measure $CO_2$ to the ground, we initially screen all thick and thin cloud cases from processing and use only clear sky cases for these results. Although it's technically possible to also include some thin cloud results, the SNR in those cases is significantly lower than for surface returns without interference from thin clouds or other scatterers, and we do not include them in the current analysis. Ranging capability discussed in previous sections is also critical for low cloud and ground discrimination. We currently distinguish between ground returns and cloud returns by comparing the distance we measure with known elevation data. If the difference in range to the surface is observed to be more than 100 meters, it is assumed to be due to the intervening presence of a thick cloud. If the range comparison is within 100 m, it is assumed to be a ground return. The presence of multiple scatterers indicates the presence of thin clouds. Note that in our archived MFLL $XCO_2$ data product (located at https://actamerica.ornl.gov/), we report these cloudy weather conditions with their corresponding cloud flags for cautions in using the data in these conditions.

## 2.7 Modeling gas absorption, spectroscopy and $XCO_2$
### 2.7.1 Gas absorption

As mentioned in previous sections, modelled DAOD values at an assumed gas concentration are needed in retrieving column gas concentrations from DAOD measurements. Optical depth for the kth species may be calculated from,

$$\tau_k = \int_{h_{surf}}^{h_{max}} X_k D A_v \sigma_k \, dz \tag{3}$$

where $X_k$ is the moist molar mixing ratio for the absorbing species, $D$ is the molar density, $A_v$ is Avogadro's number, and $\sigma_k$ is the molecular absorption cross section for the kth species. By using the ideal gas law, the hydrostatic equation, and expressing the water vapor mixing ratio in terms of the specific humidity it can be shown that,

$$\tau_{CO2} = \frac{A_v R_d}{R} \int_{P_{min}}^{P_{surf}} \frac{X'_{CO2}}{g}(1-q_v)(\sigma_{on}-\sigma_{off})dP \;, \tag{4}$$

where $\tau_{CO2}$ is the DAOD for $CO_2$, $\sigma_{on}$ is the online molecular absorption cross section for the $CO_2$ online wavelength, $\sigma_{off}$ is the online molecular absorption cross section for the $CO_2$ offline wavelength, $R_d$ is the specific gas constant for dry air, $q_v$ is the specific humidity, $X'_{CO2}$ is the dry molar mixing ratio for $CO_2$, g is the acceleration of gravity. For water vapor the optical depth is given by,

$$\tau_{H2O} = \frac{A_v R_v}{R} \int_{P_{min}}^{P_{surf}} \frac{q_v}{g}(\sigma'_{on}-\sigma'_{off})dP, \tag{5}$$

where $\tau_{H2O}$ is the DAOD for $H_2O$, , The total DAOD in this model will be the sum of the two optical depths for $CO_2$ and $H_2O$, $\sigma'_{on}$ is the molecular absorption cross section for water vapor online wavelength, and $\sigma'_{off}$ is the molecular absorption cross section for water vapor offline wavelength. There may be a small insignificant contribution from other species at our wavelengths. This form is particularly useful because MERRA-2 reports specific humidity as one of the data products and pressure is the independent variable.

From the above we see the dry column $CO_2$ must be

$$\text{XCO}_2 = \frac{\frac{A_v R_d}{R} \int_{P_{\min}}^{P_{\max}} \frac{X'_{\text{CO2}}}{g}(1-q_v)(\sigma_{on} - \sigma_{off}) dP}{\frac{A_v R_d}{R} \int_{P_{\min}}^{P_{\max}} \frac{1}{g}(1-q_v)(\sigma_{on} - \sigma_{off}) dP} = \frac{\int_{P_{\min}}^{P_{\max}} X'_{\text{CO2}}(P) W(P) dP}{\int_{P_{\min}}^{P_{\max}} W(P) dP}, \tag{6}$$

where

$$W(P) = \frac{A_v R_d}{R g(P)}(1 - q_v(P))(\sigma_{on}(P) - \sigma_{off}(P)) \tag{7}$$

is the dry air column weighting function. The uncertainty in XCO$_2$ due to an uncertainty in pressure assuming a uniform mixing ratio is given by (Dufour et al., 2003),

$$\frac{\delta \text{XCO}_2}{\text{XCO}_2} \approx \frac{W(P)}{\int_{P_{\min}}^{P_{surf}} W(P) dP} \delta P = \overline{W}(P) \frac{\delta P}{P}, \tag{8}$$

where $\overline{W}$ is the normalized weighting function. This is the absolute limit in error due to an error in pressure or range measurement assuming a perfect instrument and no other measurement errors. Near the surface $\overline{W}(P) \sim 1$, so a 1 mb uncertainty in surface pressure leads to about a 0.4 ppm uncertainty in XCO$_2$ for a single measurement. If the uncertainty is random this may be improved further through averaging.

In this model we need to take into account the variation of gravity with altitude and location. The acceleration of gravity at the surface may be calculated using the Somigliana Gravity Formula (van Dam et al., 2010),

$$g_0(\phi) = g_e \frac{1 + p \sin^2(\phi)}{\sqrt{1 - e^2 \sin^2(\phi)}}, \tag{9}$$

where $\phi$ is the latitude, $p = 1.931851353 \times 10^{-3}$, $e^2 = 6.69438002290 \times 10^{-3}$, $g_e = g_0(0) = 9.780318 \text{ m/s}^2$. A height correction may be found by [Wenzel, 1989],

$$g - g_0(\phi) \approx \left(3.0877 \times 10^{-6} / \text{s}^2 - (4.3 \times 10^{-9} / \text{s}^2) \sin^2(\phi)\right) h + \left(7.2 \times 10^{-13} / \text{ms}^2\right) h^2. \tag{10}$$

### 2.7.2 Spectroscopy

In calculating gas absorption cross section, a spectroscopic model is used. This spectroscopic model is based on a pre-release version of HITRAN2016 (Gordon et al., 2017) with inclusion of the line-mixing model of Lamouroux et al. (2015). Calculations were converted to a custom ACOS/OCO-2 ABSorption COefficient (ABSCO) lookup table (Payne and Thompson, 2013). Note that the intensities of the strongest CO$_2$ lines in the 30012-00001 band (which is the strongest absorber in this spectral region) in that version of the database were identical to those in HITRAN2012 (Rothman et al., 2013) and originate from the empirically-derived data (Toth *et al.*, 2006). The official release of the HITRAN2016 database featured ab initio intensities (Zak et al., 2016) for almost all bands of all isotopologues of $CO_2$ in all spectral regions. Comparisons with recent experimental data in other near-infrared regions have shown that the ab initio data are superior to that of Toth et al. (2006). See for instance Figure 7 of the HITRAN2016 paper (Gordon et al., 2017) for the evaluation of the 20012-0001 band at 2 μm. The ab initio intensities of the 30012-00001 band targeted here are systematically larger than those of Toth et al. (2006) by 2%. At the moment it is still arguable if this change (in this particular band) is justified, as alternative ab initio calculations (Huang et al., 2014, 2017) favor Toth et al. (2006) data while independent experiments cannot provide conclusive justification (see for instance Devi et al., 1998; Jacquemart et al., 2012) at the required level of accuracy. New experiments and atmospheric validations are now underway to resolve this issue (c.f., Fleisher et al., 2019), and we may have to reevaluate our model to accommodate data from the official HITRAN2016 release.

### 2.7.3 Geopotential and geometric heights

Except for the calibration where meteorological vertical profiles can be derived from aircraft spiral measurements, all weather information along flight tracks is obtained from MERRA-2. The MERRA data contains profiles for temperature, pressure, geopotential height, and specific humidity which are stored as four-dimensional arrays and are functions of time and location. It also includes surface pressure, sea level pressure, and geopotential surface height as a function of time and location. The hypsometric pressure and adiabatic temperature models are used to fill in missing profile data where necessary. All geopotential heights are converted to geometric heights. For geopotential height, a particularly useful closed form approximation discovered by M. J. Mahoney comes in the form where the geopotential height is given by (van Dam et al., 2010),

$$Z \approx \frac{g_0(\phi)}{g_0(45°)} \frac{h}{1 + h/R_e(\phi)}, \quad (11)$$

where Z is the geopotential height, h is the geometric height,

$$R_e(\phi) = \frac{1}{\sqrt{\frac{\cos^2(\phi)}{a^2} + \frac{\sin^2(\phi)}{b^2}}}, \quad (12)$$

$a = R_e(0°) = 6378137.0\,\mathrm{m}$, and $b = R_e(90°) = 6356752.3\,\mathrm{m}$. Solving this expression for geometric height gives,

$$h = \frac{g_0(45°)}{g_0(\phi)} \frac{Z}{1 - (g_0(45°)/g_0(\phi)) Z / R_e(\phi)}. \quad (13)$$

### 2.7.4 XCO2 from DAOD measurements

For simplicity we model the measured DAOD as

$$\tau_{meas} \simeq \tau_{CO2} + \tau_{H20} \quad (14)$$

To calculate XCO$_2$ from measured optical depth and MERRA-2 weather profiles we therefore use

$$\text{XCO}_2 = \frac{\tau_{meas} - \tau_{H20}}{\tau'_{CO2}} 400 \, ppm, \quad (15)$$

where $\tau'_{CO2}$ is calculated assuming a constant 400 ppm in Eq. 4.

## 2.8 Altitude dependent bias correction

Small differences were found as a function of range between modeled optical depths, as determined from Eq.4 and Eq.5, and the *in situ* CO$_2$ mole fraction measurements, and MFLL measured optical depths, which are determined from Eq. 2. There are multiple potential reasons causing these differences, such as the previously discussed uncertainties in the pressure-, temperature-, and humidity-dependent spectroscopic model, unknown cross-talks between online and offline channels, non-linear response of the instrument, and/or spatiotemporal differences between in situ profiling measurements and lidar overpasses. As a result, for each ACT-America campaign, we determined what the altitude-dependent bias correction needed to be to make the MFLL optical depths best agree with the model results. To determine the altitude bias correction, we collected all the available coincident lidar and *in situ* data. At multiple times and locations throughout each campaign, we collect MFLL observations over a cloud-free location where we also obtained an *in situ* profiles of CO$_2$ mole fraction from the altitude of the MFLL data collection to approximately 300 m AGL (above ground level) or lower in some cases where airport missed approaches were arranged. All flights are conducted in relatively well-mixed, daytime conditions to minimize near-surface gradients. Because the retrieval is based on Merra-2 meteorology, we then calculated the model optical depths from the Merra-2 meteorological and *in situ* CO$_2$ profile measurements based on Eq. 4 and Eq. 5. The advantage of using Merra-2 in the calibration is whatever repeatable errors exist in Merra-2, can be calibrated out to a certain degree because the retrieval is based on the same meteorology. We then compared the model optical depths to the measured optical depths. Since we already subtract off the instrument optical depth and range calibration offsets from the ground testing, as discussed in section 2.3, we assume a relationship for the fractional change as

$$(\tau_{meas} - \tau_{mod})/\tau_{meas} = k_1 + k_2 \tau_{meas}. \quad (16)$$

To find $k_1$ and $k_2$ we minimize variance with respect to $k_1$ and $k_2$ which is given by

$$\sum_i \left((\tau_{meas,i} - \tau_{mod,i})/\tau_{meas,i} - k_1 - k_2 \tau_{meas,i}\right)^2. \quad (17)$$

Once $k_1$ and $k_2$ are found the final correction is determined from

$$\tau' = \tau_{meas} - (k_1 + k_2\tau_{meas})\tau_{meas} = (1-k_1)\tau_{meas} - k_2\tau_{meas}^2. \qquad (18)$$

## 3. Results

### 3.1 Bias correction

As an example, we take multiple altitude cases from spiral overpasses collected from the 2017-2018 campaigns where each altitude case represents one data point. The data points used are calculated from the left hand side of Eq.16 by taking the difference between the measured optical depths and the modeled optical depths calculated from Eq. 4 and Eq. 5 using Merra-2 pressure, temperature, humidity, and in situ $CO_2$ measurements. In the case of $CO_2$, the airborne Picarro measurements were used. Once the *in situ* measurements were obtained, the modeled optical depths were calculated from our spectroscopy model. In this particular case we found $k_1$= 0.01057 and $k_2$= -0.04304 (Spring 2018), $k_1$= 0.01035 and $k_2$= -0.03979 (Fall 2017), and $k_1$= 0.01063 and $k_2$= -0.04114 (Winter 2017). A plot of these results are shown in Figure 2a. The solid lines are the fits defined by Eq. 16 and the dots are the data. Figure 2b was computed using $\delta XCO_2 \approx 400\,\delta\tau/\tau_{meas}$, where $\delta\tau$ is the difference between the measured points and the straight line fits. Conversely, we can estimate the maximum change due to the bias correction as $\delta XCO_2 \approx 400(k_1 + k_2\tau_{max})$. In each case the maximum change is approximately 4 ppm or less. The maximum difference between each season can be computed by finding the difference of the maximum change for each season. This is about 0.5 ppm.

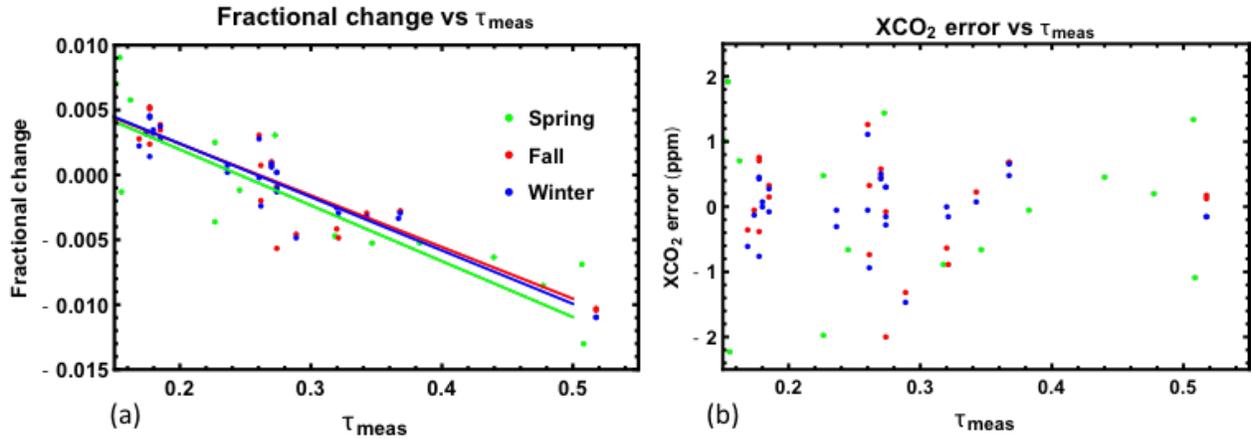

Figure 2. Plot of fractional change of optical depth vs measured optical depth (a) and residual $XCO_2$ error in ppm vs measured optical depth (b).

A plot of the residual $XCO_2$ errors in ppm are shown in Figure 2b. In this case the mean is zero with a standard deviation of 0.8 ppm. When targets were close to aircraft, there would be small atmospheric gas absorption. Thus, lidar absorption signals would be weak, and retrieved DAOD values as shown in Fig. 2b would be noisier than for higher DAODs. This may reflect the absolute accuracy caused by a combination of multiple uncertainty sources as discussed previously. Once the bias calibration of Eq. 18 is applied to MFLL measurements, the final $XCO_2$ is estimated from the scaling of measured $CO_2$ DAOD and model DAOD based on meteorological state profile data with assumed 400 ppm $CO_2$ (c.f., Eq. 15). The most encouraging result is the consistency between altitude-dependent calibration results going from season to season and in different measurement environments. This consistency gives some indication of the long term stability of the instrument and shows a potential that if it is calibrated in one field campaign, that the calibration more or less holds for future measurements. By calculating the difference between the straight line fits in Fig. 2, we found the maximum equivalent difference between each calibration is less than 0.5 ppm.

**3.2 In-flight testing of measurement precision and stability**

To fully understand instrument performance, besides absolute accuracy, other key parameters such as measurement precision, as reflected by measurement Signal-to-Noise Ratio (SNR), and in-flight instrument stability over hours need to be evaluated. Before discussing this it should be pointed out this topic has been addressed in a previous study (Lin et al., 2013), where specific instrument parameters were modeled and precision and accuracy were calculated. For this study, dedicated MFLL flight experiments were conducted. The selected region was the Gulf of Mexico, which was chosen due to the relatively homogeneous conditions there during on-shore flow. On 6 Nov 2017, the synoptic set up over the southern portion of the Gulf States and over the Gulf of Mexico was characterized with a broad high-pressure system with moderate southerly wind in the boundary layer. The flight plans for both C-130 and B-200 aircraft (Fig. 3a) were designed to examine the performance of the MFLL system under homogeneous atmospheric condition over water. The B-200 aircraft made two west-east Atmospheric Boundary Layer (ABL) legs at an altitude of 400 m above MSL (Mean Sea Level) while the C-130 made five west-east transacts in the Free Troposphere (FT) at an altitude of 4800 m above MSL; additionally five vertical profiles were made over water using both aircraft which confirmed a spatially homogeneous boundary layer depth of 700 m above MSL over water (not shown here).

Figure 3b displays a longitude-versus-altitude cross section of the in situ measurments of atmospheric $CO_2$. The column averaged $CO_2$ mole fraction obtained from in situ measurements was 404.0 ppm. In Fig. 3b, two ABL legs are stacked on each other (same latitude); thus, we seperated the two ABL legs in Fig.3c where we present a time-series view of $CO_2$ variability from west-to-east within the ABL. These data confirm that $CO_2$ distributions along ABL-Leg 1 and ABL-Leg 2 were similar, thus atmospheric $CO_2$ was steady with respect to time since ABL-Leg 2 was flown more than an hour after ABL-Leg 1. Figure 3c illustates that the $CO_2$ horizontal distribution in the boundary layer over the water was also spatially homogeneous during both transacts. The mean, median, and standard deviation of $CO_2$ measurements along ABL-Leg 1 (ABL-Leg 2) were 405.15 ppm (405.11 ppm), 405.17 ppm (405.29 ppm), and 0.09 ppm (0.08 ppm), respectively. These in situ observations demostrate that although there was some vertical variation, the $CO_2$ variability at sampled constant altitudes was minimal, even within the atmospheric boundary layer. Thus, spatiotemporal $XCO_2$ variations must be small for this case, which is what was desired for our precision and stability testing.

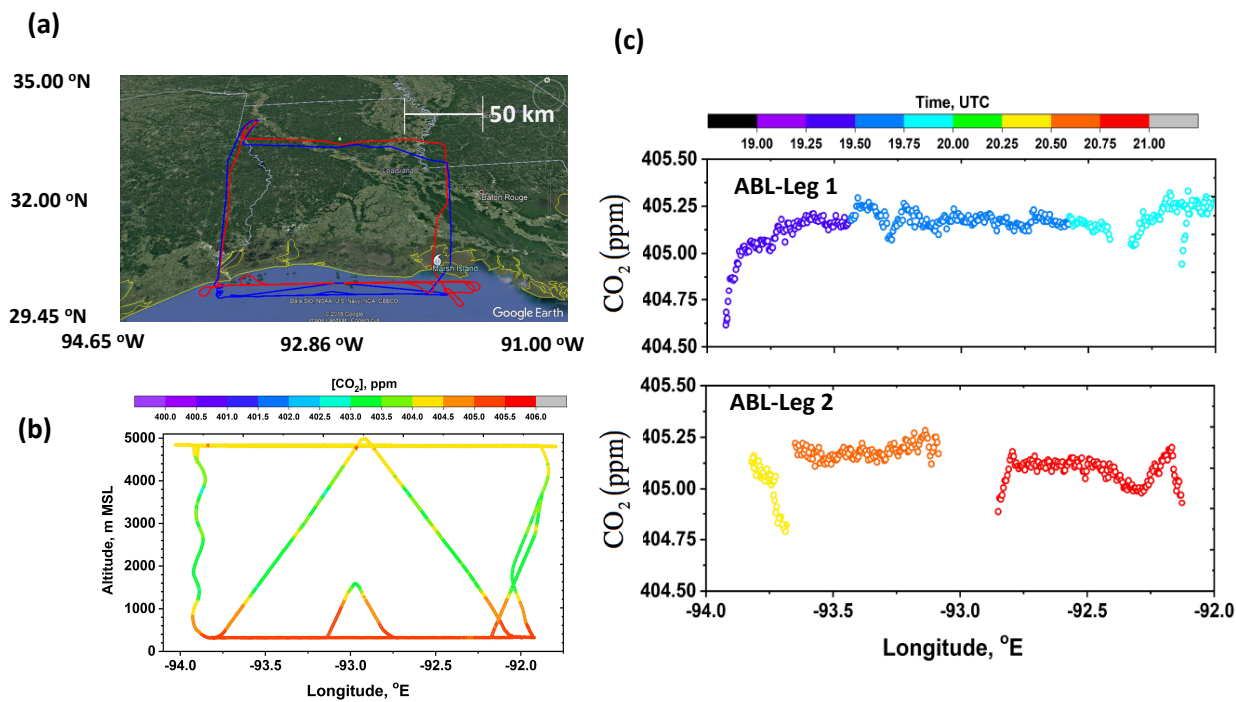

Figure 3. Ground tracks of B-200 (blue) and C-130 (red) aircraft on November 6, 2017 during ACT-America Fall field campaign (a). Longitude-versus-altitude cross-section of 5-s averaged samples of $CO_2$ (ppm) obtained by Picarro onboard both C-130 and B-200 (b). Time-series view of east-west transacts of $CO_2$ in the boundary layer over Gulf of Mexico along two identical tracks: Leg 1 (upper panel) and Leg 2 (lower panel) (c)

Figure 4 shows the 1-s, 10-s and 60-s running mean variations of the $XCO_2$ retrieved from the MFLL measurements. The native instrument measurement rate is 10 Hz, and the values of $XCO_2$ retrievals at this 10-Hz reporting rate is dominated by instrument noise and needed to be further averaged before science use.

Although the $CO_2$ lidar system maintained continuous operation during the entire flight period, the plotted data showed some gaps that resulted from multiple reasons. Removing remotely sensed data from retrievals due to aircraft turns over the Gulf of Mexico was one. Others were caused during the data processing such as screening of large aircraft attitude angles and/or cloudy scenes, as mentioned previously. It can be seen that with the longer averaging times, the random errors reduced significantly. Generally, when the averaging time was longer than 10 s. the variation in retrieved $XCO_2$ was within 1 ppm.

The estimated precision (standard deviation) for 0.1-s, 1-s, 10-s, and 60-s averages were determined to be 3.4 ppm (0.84%), 1.2 ppm (0.30%), 0.43 ppm (0.10%), and 0.26 ppm (0.063%), respectively. Their corresponding SNRs, which are the mean $XCO_2$ divided by the standard deviation, were found to be 120, 330, 950, and 1600, respectively. The 10-s results are very close to previously reported MFLL results (Dobler et al. 2013) after surface type and flight altitude differences are taken into consideration and consistent with that predicted in a previous study (Lin et al., 2013).

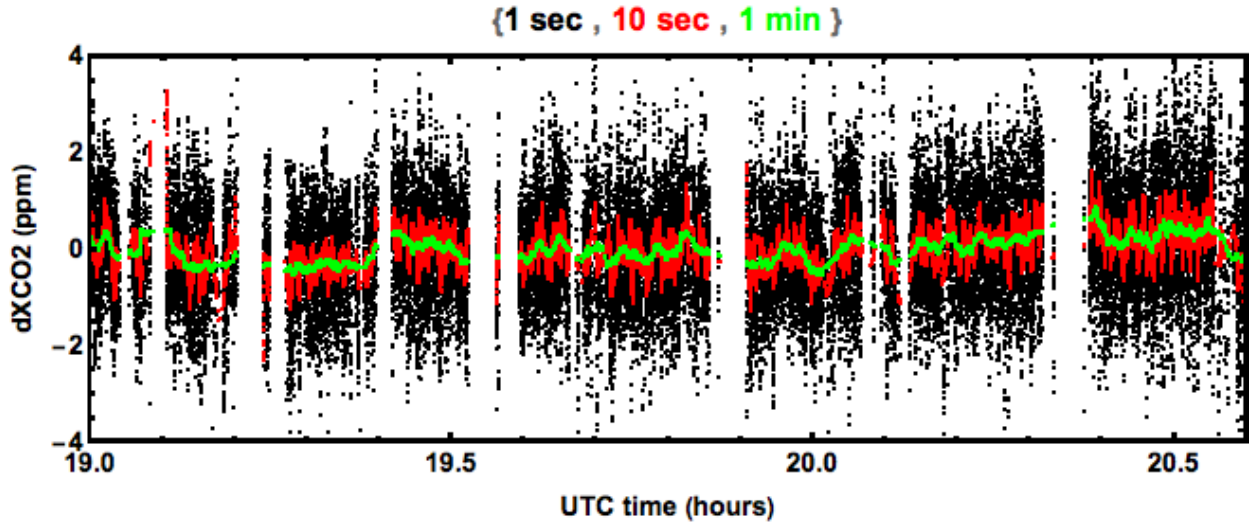

Figure 4. Variations of the retrieved XCO$_2$ for the Gulf of Mexico flight on November 6, 2017. Each segment represents a repeated pass over the same segment of the atmosphere. Here the 1-s, 10-s and 60-s running means, as indicated by their corresponding color, are shown.

Column XCO$_2$ derived from MFLL (Figure 4) shows very little change during the five repeated passes over this portion of the Gulf of Mexico, consistent with the homogeneous, steady-state atmospheric conditions described in Figure 3. For the in-flight instrument stability, the trends at 20-minute, 30-minute, and one hour scales in the 10-s averaged Gulf data were analyzed. No statistically significant drifts were found.

This flight duration over the Gulf of Mexico was generally very similar to that of OCO-2 under-flights. However, OCO-2 under flights were conducted at about 8.5 km above MSL altitudes over land. Since land surface lidar reflectance is about 4 times greater than over water (Dobler et al., 2013), lidar return signals from the sea surface at this 4.8 km altitude would be weaker than those for OCO-2 under flights. Other ACT-America flights were equal to or lower than those of OCO-2 under flights and nearly all were over land. Thus, lidar signal level from this experiment would be among the weakest for ACT-America flight campaigns. The estimated precision and SNR would be a conservative result for normal ACT-America instrument operations.

We additionally evaluate instrument precision from an 8.4 km altitude OCO-2 under-flight conducted across the US Great Plains on October 22, 2017. The flight track and MFLL observed XCO$_2$ are shown in Figs. 5a and 5b, respectively. In this case some of the observed XCO$_2$ variations were caused by atmospheric differences in CO$_2$ concentrations, variability in meteorological state variables, and boundary layer thickness and surface height variations. The estimated precision for 0.1-s, 1-s, 10-s, and 60-s averages were calculated as 4.6 ppm (1.1%), 1.5 ppm (0.37%), 0.82 ppm (0.20%), and 0.46 ppm (0.11%), respectively. Their corresponding SNRs were 88, 270, 490, and 909, respectively. To facilitate the validation of remote-sensing measurements with in situ data, NASA's Global Modeling and Assimilation Office (GMAO) assimilates the in situ data into the Goddard Earth Observing System (GEOS). The result, a "gap-filled", two-dimensional, along-track cross-section, referred to here as a curtain, is then directly comparable to the remote-sensing measurements taken on the same platform. To produce the curtains, the assimilation system uses a model configuration similar to that described in Ott et al. (2015) and an analysis methodology derived from that used to produce the three-dimensional fields of OCO-2 data highlighted in Eldering et al. (2017). For more details about the relatively minor differences from the system described in those publications, see Bell et al. (2019). The general consistency of these MFLL $XCO_2$ measurements when compared to GEOS assimilated results and OCO-2 observations (not shown here, c.f., Bell et al., 2019) further demonstrate the stability and precision of the MFLL $XCO_2$ measurements.

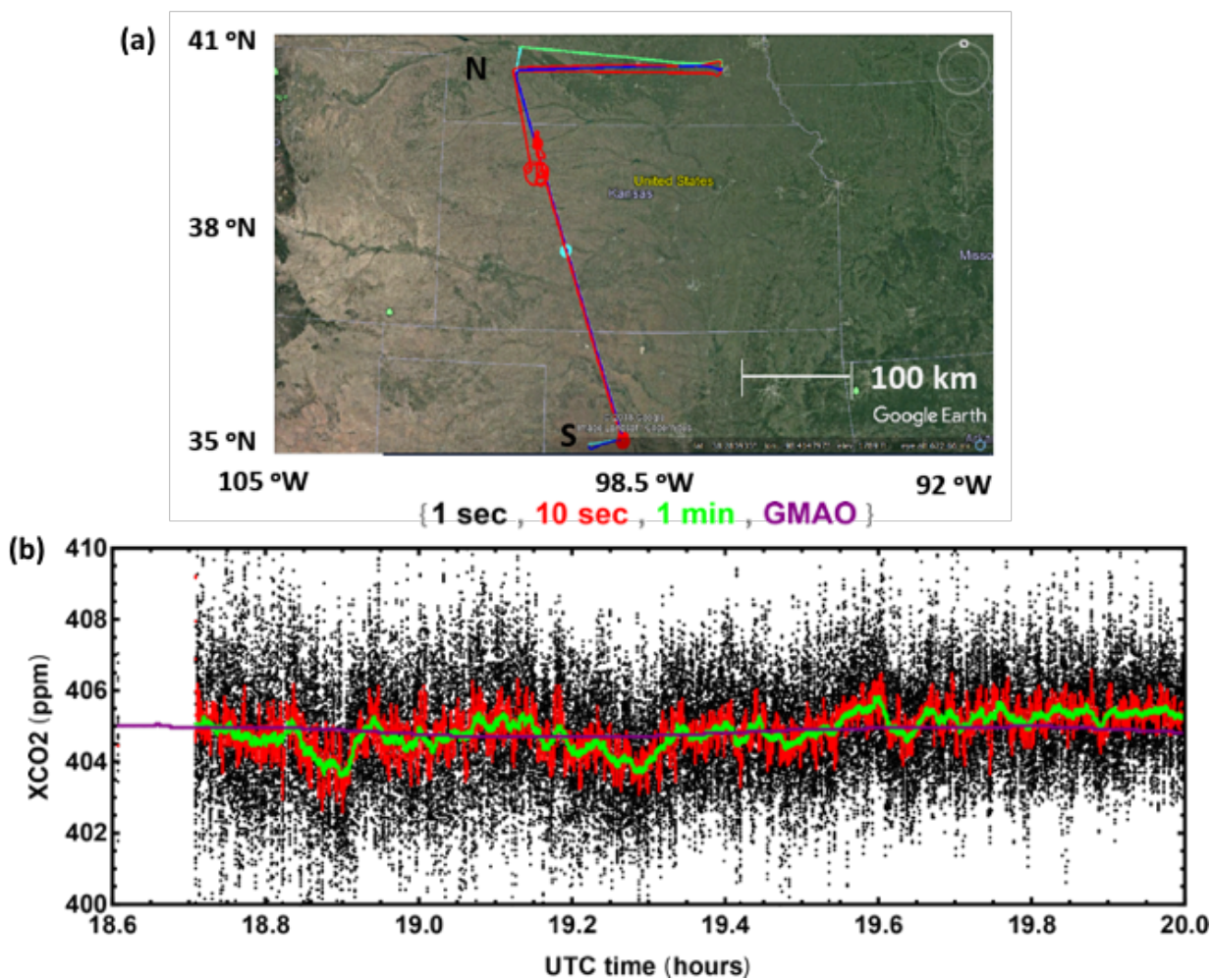

Figure 5. Flight track (a) and variations of the retrieved $XCO_2$ (b) for the land case across the US Great Plains on October 22, 2017. Each segment represents a repeated pass over the same segment of the atmosphere. Here the 1-s, 10-s and 60-s running means, as indicated by their corresponding color, are shown, which were compared with GMAO data (purple).

For both the Gulf and land cases and with 10-s and longer averaging times, the MFLL $XCO_2$ measurement precision clearly meets the ACT-America science requirement, which is 1 ppm at 20-km spatial scale corresponding to about 2.6 minutes of C-130 flight time. Even the maximum deviation of about 0.8 ppm, as shown in the green curve in Fig. 4 for 60-s averaged data, well below 1 ppm.

## 4. Conclusion

We have demonstrated a technique for measuring $XCO_2$ with an IM-CW lidar system through flight-testing with very good results. The lidar measurement precision for 1, 10 and 60-second averages (approximately 0.13, 1.3 and 8 km) is between 0.9-1.2 ppm, 0.4 – 0.6 ppm, and 0.2 – 0.4 ppm, respectively, and on one-hour time scales, the average drift is about 0.07 ppm, which is small compared to typical atmospheric variations in atmospheric $CO_2$. These results demonstrate that the precision and stability of our $CO_2$ lidar measurements meet ACT-America science requirements. Furthermore, our technique is suitable as a standard tool for remotely measuring atmospheric $CO_2$ column concentrations with an airborne lidar system that has been in development and testing for many years. We found our altitude-dependent calibration holds over many seasons which is a major discovery because it means that we can calibrate it using in situ profiling observations as a reference in one field campaign and have reasonable expectation the same calibration applies for future campaigns without the need for extensive recalibration flight experiments. We found our $XCO_2$ measurement accuracy was about 0.8 ppm compared against our model results derived from in situ CO2 measurements based on the state-of-the-art HITRAN spectroscopic model and Merra-2 meteorology. By using meteorology consistent with the retrieval in the calibration we can take out potential bias errors in the derived XCO2 retreval caused by bias errors in Merra-2. There is still an unanswered question regarding the cause(s) of the altitude-dependent bias errors even though we have demonstrated we can calibrate for it with excellent results. In the future we plan to investigate this question and apply the MFLL $XCO_2$ measurements in studying $CO_2$ variability in different atmospheric conditions being investigated in ACT-America.

**Acknowledgements**
ACT-America data supporting this work can be found at https://www-air.larc.nasa.gov/missions/ACT-America/ and at https://doi.org/10.3334/ORNLDAAC/1649. Merra-2 meteorological data can be found at https://gmao.gsfc.nasa.gov/reanalysis/MERRA-2/data_access/. Co-author Sandip Pal was supported by NASA Grant Number 80NSSC19K0730 and Texas Tech University start up research grant 14A001-B5399P-200.